\documentclass[12pt,thmsa]{article}

\def \eq {\begin{equation}}
\def \fim-eq {\end{equation}}
\begin{document}

\author{E. S. Guerra \\
Departamento de F\'{\i}sica \\
Universidade Federal Rural do Rio de Janeiro \\
Cx. Postal 23851, 23890-000 Serop\'edica, RJ, Brazil \\
email:emerson@ufrrj.br\\
and \\
C. R. Carvalho \\
Instituto de F\'{\i}sica \\
Universidade Federal do Rio de Janeiro \\
Cx.Postal 68528, 21945-970 Rio de Janeiro, RJ, Brazil \\
email:crenato@if.ufrj.br}
\title{ENTANGLEMENT SWAPPING: ENTANGLING ATOMS THAT NEVER INTERACTED}
\maketitle

\begin{abstract}
\noindent In this paper we discuss four different proposals of entangling atomic states of particles which have never interacted.
The experimental realization proposed makes use of the interaction of
Rydberg atoms with a micromaser cavity prepared in either a coherent state
or in a superposition of the field Fock states $|0\rangle$ and $|1\rangle$.
We consider atoms in either a three-level cascade or lambda configuration.

\ \newline

\end{abstract}

\section{\protect\bigskip INTRODUCTION}

One of the most intriguing aspects, as well as fundamental, in quantum
mechanics is entanglement. This feature was first noticed by Einstein,
Podolsky and Rosen \cite{epr} who have originally proposed the EPR
experiment in order to show that quantum mechanics were not a complete
theory to describe reality. At the same time Schr\"odinger has done a formal
discussion about the description and the measurements performed on two system which have interacted and that are far apart from each other 
\cite{schrodinger}. Entanglement is of central importance in Bell's theorem and is the origin of nonlocality in quantum mechanics. Usually
entanglement is understood as a consequence of some interaction of the
particles in their common past. Thus far, it has been achieved either by
having the two particles emerging from the same source \cite{freedman} or by
having the two particles interacting with each other \cite{lamehi}. However,
Yurke and Stoler \cite{yurke} and Zukowski \textit{et al} \cite{zukowsky}
showed that one can entangle particles that do not even share any common
past. The realization of the entanglement swapping was done by Pan 
\textit{et al} \cite{pan} in a scheme involving the horizontally and
vertically polarized states of photons.

In a recent paper it has been suggested an experimental realization of
teleportation of atomic states via cavity quantum eletrodynamics \cite
{guerra01}. In this work, Guerra discusses theoretically how to prepare and
how to detect atomic Bell states. Based on the results presented there, in the present work we show that it is possible to build a scheme  to entangle the states of two atoms which have never interacted, along a similar line of Ref.\cite{pan}. In section 2 we review the process of preparing
atomic Bell states for different atomic configurations and cavity fields. In
section 3 we show that, through appropriate measurements of the Bell states, we can perform the entanglement swapping and, finally, in section 4 we discuss our conclusions.

\section{REALIZATION OF BELL STATES}

In this section, we present four different schemes to prepare Bell
states involving two different configurations of the atomic levels.

\subsection{Cascade atomic system and cavity in a coherent state}

\label{casc-coer}

Consider a three-level cascade atom \ $A_k$ with $|e_{k}\rangle ,
|f_{k}\rangle $ and $| g_{k}\rangle $ being the upper, intermediate and
lower atomic states (see Fig. 1). We assume that the transition $|
e_{k}\rangle \rightleftharpoons | f_{k}\rangle $ is far enough from
resonance with the cavity frequency. Consequently, it
can be shown, according to the Jaynes-Cummings model \cite{Orszag}, that
this dispersive interaction can be represented by the time evolution
operator 
\begin{equation}
U=e^{-i\varphi (a^{\dagger }a+1)} |e_k\rangle \langle e_k| +e^{i\varphi
a^{\dagger }a}| f_k\rangle \langle f_k| ,  \label{U-ef}
\end{equation}
where $a$ $(a^{\dagger })$ is the annihilation (creation) operator for the
field in cavity $C$, $\varphi =g^{2} \tau /$ $\Delta $, \ $g$ is the
coupling constant, $\tau $ is the interaction time, $\Delta =\omega
_{e}-\omega _{f}-\omega $ is the detuning, $\omega _{e}$ and $\omega _{f}$ \
are the frequencies of the upper and intermediate levels respectively and $
\omega $ is the cavity field frequency.

In addition we assume that the transitions $| f_{k}\rangle
\rightleftharpoons | g_{k}\rangle $ and $| e_{k}\rangle \rightleftharpoons |
g_{k}\rangle $ are highly detuned from the cavity frequency so that there
will be no coupling with the cavity field involving the state $|
g_{k}\rangle $. However the state $| g_{k}\rangle$ plays an important role
in the following process, because we suppose that it is coupled to the state $| f_{k}\rangle $ in the Ramsey cavities which we shall use to prepare the
atomic Bell states involving the states $| f_{k}\rangle$ and $|g_{k}\rangle$.

Therefore, considering the atom-field interaction, the level 
$| e_{k}\rangle$ will never be populated during the whole process so that we can ignore it from now on, since it will not play any role
in our scheme, being important only as origin of the phase factor in the time evolution operator (see Eq.(\ref{U-ef})) due to the
dispersive interaction. Hence, we have effectively a two-level system
involving the states $| f_{k}\rangle $ and $|g_{k}\rangle $. In terms of
them we can write the time evolution operator 
\begin{equation}
U=e^{i\varphi a^{\dagger }a}| f_{k}\rangle \langle f_{k}| +|g_{k}\rangle
\langle g_{k}| ,  \label{U-fg}
\end{equation}
where the second term above was put by hand just in order to take into
account the effect of level $| g_{k}\rangle $. Let us take $\varphi =\pi $.

We assume that we have a two-level atom $A_1$ initially in the state $|
g_{1}\rangle $ which, after crossing a Ramsey cavity $R_1$, is prepared in a
coherent superposition 
\begin{equation}
| \psi \rangle _{A_1}=\frac{1}{\sqrt{2}}(| f_{1}\rangle +| g_{1}\rangle ),
\end{equation}
according to the rotation matrix 
\begin{equation}
\frac{1}{\sqrt{2}}\left[ 
\begin{array}{cc}
1 & 1 \\ 
-1 & 1%
\end{array}
\right].  \label{M-A}
\end{equation}

Now, let atom $A_1$ cross the cavity $C$ which is prepared in coherent state 
$|-\alpha \rangle $. A coherent state $|\beta \rangle $ is obtained by
applying the displacement operator $D(\beta )=e^{(\beta a^{\dag }-\beta
^{\ast }a)}$ to the vacuum, i.e. $|\beta \rangle =D(\beta )|0\rangle$ \cite%
{Louisell} and, experimentally, it is obtained with a classical oscillating
current in an antenna coupled to the cavity. Then, according to (\ref{U-fg})
the system $A_1-C$ evolves to 
\begin{equation}
| \psi \rangle _{A_1-C}=\frac{1}{\sqrt{2}}(| f_{1}\rangle |\alpha \rangle +|
g_{1}\rangle |-\alpha \rangle ),
\end{equation}
where we have used $e^{za^{\dagger }a}|\alpha \rangle =|e^{z}\alpha \rangle $
\cite{Louisell}$.$ Now, if atom $A_1$ enters a second Ramsey cavity $R_2$,
where the atomic states are rotated according to the rotation matrix (\ref%
{M-A}), we obtain 
\begin{equation}
| \psi \rangle _{A_1-C}=\frac{1}{2}\Biggl\{| f_{1}\rangle \Bigl(~|\alpha
\rangle + |-\alpha \rangle ~\Bigr)-| g_{1}\rangle \Bigl(~|\alpha \rangle
-|-\alpha \rangle ~\Bigr)\Biggr\}.
\end{equation}

Consider now a second atom $A_2$ prepared in the same way of atom $A_1$, so
that, before entering the cavity $C$, we have it in the state 
\begin{equation}
| \psi \rangle _{A_2}=\frac{1}{\sqrt{2}}(| f_{2}\rangle +| g_{2}\rangle ).
\end{equation}
Let us send this atom through cavity $C$ assuming that \ for atom $A_2$, as
above for atom $A_1$, the transition $| f_{2}\rangle \rightleftharpoons |
e_{2}\rangle $ is far from resonance with the cavity frequency. As above, taking into account the time evolution operator $U$, Eq. (\ref{U-fg}%
) with $\varphi =\pi $, after the atom has passed through the cavity, we get 
\begin{equation}
| \psi \rangle _{A_1-A_2-C}=\frac{1}{2\sqrt{2}}\Biggl\{| f_{1}\rangle \Bigl(%
~|f_{2}\rangle +| g_{2}\rangle ~\Bigr)\Bigl(~|\alpha \rangle +|-\alpha
\rangle ~\Bigr)+ | g_{1}\rangle \Bigl(~| f_{2}\rangle -| g_{2}\rangle ~\Bigr)%
\Bigl(~|\alpha \rangle -|-\alpha \rangle ~\Bigr)\Biggr\}.
\end{equation}
Then, atom $A_2$ enters a Ramsey cavity $R_3$, where the atomic states are
also rotated according to the rotation matrix (\ref{M-A}), which results in 
\begin{equation}
\frac{1}{\sqrt{2}}( |f_{2}\rangle + |g_{2}\rangle )\rightarrow |
f_{2}\rangle ~~\mbox{ and }~~ \frac{1}{\sqrt{2}}( |f_{2}\rangle -
|g_{2}\rangle )\rightarrow - | g_{2}\rangle
\end{equation}
leading to 
\begin{equation}
| \psi \rangle _{A_1-A_2-C}=\frac{1}{2}\Biggl\{| f_{1}\rangle |f_{2}\rangle %
\Bigl(~|\alpha \rangle +|-\alpha \rangle ~\Bigr)- |g_{1}\rangle
|g_{2}\rangle \Bigl(~|\alpha \rangle -|-\alpha \rangle ~\Bigr)\Biggr\}.
\label{A_1A_2C}
\end{equation}
Now, we inject $|-\alpha \rangle $ in cavity $C$ which mathematically is
represented by the operation $D(\beta )|\alpha \rangle =|\alpha +\beta
\rangle $ \cite{Louisell} and this gives us 
\begin{equation}
| \psi \rangle _{A_1-A_2-C}=\frac{1}{2}\Biggl\{|f_{1}\rangle | f_{2}\rangle %
\Bigl(~ |0\rangle +|-2\alpha \rangle ~\Bigr) -| g_{1}\rangle | g_{2}\rangle %
\Bigl(~|0\rangle -|-2\alpha \rangle ~\Bigr) \Biggr\}.
\end{equation}

In order to disentangle the atomic states of the cavity field state we now
send a two-level atom $A_3,$ resonant with the cavity, with $|f_{3}\rangle $
and $|e_{3}\rangle $ being the lower and upper levels respectively, through $%
C$. If $A_3$ is sent in the lower state $|f_{3}\rangle $, under the
Jaynes-Cummings dynamics \cite{Orszag} we know that the state $|f_{3}\rangle
|0\rangle $ does not evolve, however, the state $|f_{3}\rangle |-2\alpha
\rangle $ evolves to $|e_{3}\rangle |\chi _{e}\rangle +|f_{3}\rangle |\chi
_{f}\rangle $, where $|\chi _{f}\rangle =\sum\limits_{n}C_{n}\cos (g\tau%
\sqrt{n} )|n\rangle $ and $|\chi _{e}\rangle =-i\sum\limits_{n}C_{n+1} \sin
(g\tau\sqrt{n+1} )|n\rangle $ and $C_{n}=e^{-\frac{1}{2}|2\alpha
|^{2}}(-2\alpha )^{n}/\sqrt{n!}$. Then we have 
\begin{equation}
|\psi \rangle _{A_1-A_2-A_3-C}=\frac{1}{2}\Biggl\{|f_{1}\rangle |f_{2}\rangle 
\Bigl(|f_{3}\rangle |0\rangle + |e_{3}\rangle |\chi e\rangle +|f_{3}\rangle
|\chi _{f}\rangle \Bigr)- |g_{1}\rangle |g_{2}\rangle \Bigl(|f_{3}\rangle
|0\rangle -|e_{3}\rangle |\chi e\rangle -|f_{3}\rangle |\chi _{f}\rangle 
\Bigr)\Biggl\},
\end{equation}
and if we detect atom $A_3$ in state $|e_{3}\rangle $ finally we get the
Bell state 
\begin{equation}
| \Phi ^{+}\rangle _{A_1-A_2}=\frac{1}{\sqrt{2}}(| f_{1}\rangle |
f_{2}\rangle +| g_{1}\rangle | g_{2}\rangle ),  \label{EPRPHI+}
\end{equation}
which is an entangled state of atoms $A_1$ and $A_2$, which in principle may
be far apart from each other.

In the above disentanglement process we can choose a coherent field with a
photon-number distribution with a sharp peak at average photon number $
\langle n\rangle =|\alpha |^{2}$ so that, to a good approximation, $|\chi
_{f}\rangle \cong C_{\overline{n}}\cos (\sqrt{\overline{n}}g\tau )|\overline{%
n}\rangle $ and $|\chi _{e}\rangle \cong C_{\overline{n}}\sin (\sqrt{%
\overline{n}}g\tau )|\overline{n}\rangle $, where $\overline{n}$ is the
integer nearest $\langle n\rangle $, and we could choose, for instance $%
\sqrt{\overline{n}}g\tau =\pi /2$, so that we would have $|\chi_{e}\rangle
\cong C_{\overline{n}}|\overline{n}\rangle $ and $|\chi _{f}\rangle \cong 0$. In this case the atom $A_3$ \ would be detected in state $|e_{3}\rangle $ with almost $100\%$ of probability. Therefore, proceeding this way, the atomic and field states will be disentangled successfully
as we would like.

Notice that starting from (\ref{A_1A_2C}) if we had injected $|\alpha
\rangle $ in the cavity and detected $|e_{3}\rangle $ we would get the Bell
state 
\begin{equation}
| \Phi ^{-}\rangle _{A_1-A_2}=\frac{1}{\sqrt{2}}(| f_{1}\rangle |
f_{2}\rangle -| g_{1}\rangle | g_{2}\rangle ).  \label{EPRPHI-}
\end{equation}

Now, if we apply an extra rotation on the states of atom $A_2$ in (\ref%
{EPRPHI+}) in a Ramsey cavity $R_4,$ according to the rotation matrix 
\begin{equation}
R_{4}=| f_{2}\rangle \langle g_{2}|-| g_{2}\rangle \langle f_{2}|,
\label{EPRR5}
\end{equation}
we get 
\begin{equation}
| \Psi ^{-}\rangle _{A_1-A_2}=\frac{1}{\sqrt{2}}(| f_{1}\rangle |
g_{2}\rangle -| g_{1}\rangle | f_{2}\rangle ),  \label{EPRPSI-}
\end{equation}
and applying (\ref{EPRR5}) on (\ref{EPRPSI-}), it yields
\begin{equation}
| \Psi ^{+}\rangle _{A_1-A_2}=\frac{1}{\sqrt{2}}(| f_{1}\rangle |
g_{2}\rangle +| g_{1}\rangle | f_{2}\rangle ).  \label{EPRPSI+}
\end{equation}
The states (\ref{EPRPHI+}), (\ref{EPRPHI-}), (\ref{EPRPSI-}) and (\ref%
{EPRPSI+}) form a Bell basis \cite{ Nielsen, BELLbasis} which are a complete
orthonormal basis for atoms $A_1$ and $A_2$.

It is important to notice that the efficacy of the process described
above depends on a measurement which has a probabilistic character: the
measurement of atom $A_3$ in the state $|e_3\rangle$ after passing the
cavity $C$. Although we can argue that the cavity $C$ may be prepared in a
coherent state $|\alpha\rangle$ with a large mean photon number $\langle
n\rangle$ so that, to a good approximation, we have almost $100\%$ of
probability of measuring atom $A_3$ in the state $|e_3\rangle$, we still have a small probability that it will fail. In the next section we
discuss an alternative way to avoid this aspect.

\subsection{Cascade atomic system and cavity in a superposition of Fock
states}

Now, we start assuming that we have the cavity $C$ prepared in the state 
\begin{equation}
|+ \rangle _{FS}=\frac{(|0\rangle +|1\rangle )}{\sqrt{2}},  \label{PsiC}
\end{equation}
where $FS$ refers to Fock state.
This state can be prepared by sending a two-level atom $A_0$, in the lower
state $|f_{0}\rangle$, first through a Ramsey cavity $R_0$, where the atomic
states are rotated according to 
\begin{equation}
\frac{1}{\sqrt{2}}\left[ 
\begin{array}{cc}
1 & i \\ 
i & 1%
\end{array}
\right] ,  \label{rot_0}
\end{equation}
and then through the cavity $C$, initially in the vacuum state $|0\rangle$.
The effect of $R_0$ is to perform the atomic state rotation 
\begin{equation}
|f_{0}\rangle \rightarrow \frac{1}{\sqrt{2}}(|f_{0}\rangle + i|e_{0}
\rangle).
\end{equation}
We assume that $A_0$ is resonant with $C$  and that $g\tau=\pi /2$. Thus, when $A_0$ cross the cavity, according to the Jaynes-Cummings model, $|f_{0}\rangle |0\rangle $ does not change whereas $|e_{0}\rangle |0\rangle $ evolves to $-i|f_{0}\rangle |1\rangle $. Hence, the passage of $A_0$ through 
$C$ yields 
\begin{equation}
\frac{(|f_{0}\rangle +i|e_{0}\rangle )}{\sqrt{2}}|0\rangle \longrightarrow
|f_{0}\rangle \frac{(|0\rangle +|1\rangle )}{\sqrt{2}}=|f_{0}\rangle
|+\rangle _{FS}.
\end{equation}
Therefore, we have prepared the cavity $C$ in the state (\ref{PsiC}).

As done in the previous section, let us consider three-level cascade atoms \ 
$A_k$ with $|e_{k}\rangle , |f_{k}\rangle $ and $|g_{k}\rangle $ being the
upper, intermediate and lower atomic state, such that the transition $%
|f_{k}\rangle \rightleftharpoons |e_{k}\rangle $ is far enough from
resonance with the cavity and the transitions $|e_{k}\rangle
\rightleftharpoons |g_{k}\rangle $ and $|f_{k}\rangle
\rightleftharpoons |g_{k}\rangle $ are highly detuned so that there will be
no coupling with the cavity field. Consider that atom $A_1$ prepared in the state 
\begin{equation}
| \psi \rangle _{A_1}=\frac{1}{\sqrt{2}}(| f_{1}\rangle +| g_{1}\rangle),
\end{equation}
interacts with cavity $C$ prepared in the state (\ref{PsiC}). Taking into
account (\ref{U-fg}), after atom $A_1$ has passed through the cavity, we get 
\begin{equation}
| \psi \rangle _{A_1-C}=\frac{1}{2}\Bigl\{(~|f_{1}\rangle + |g_{1}\rangle
~)|0\rangle +(-|f_{1}\rangle + |g_{1}\rangle ~)|1\rangle \Bigr\}.
\end{equation}
Now, if atom $A_1$ enters a second Ramsey cavity $R_2$ where the atomic
states are rotated according to the rotation matrix (\ref{M-A}), we have 
\begin{eqnarray}
& &\frac{1}{\sqrt{2}}( |f_{1}\rangle + |g_{1}\rangle )\rightarrow |
f_{1}\rangle ,  \nonumber \\
& &\frac{1}{\sqrt{2}}(- |f_{1}\rangle + | g_{1}\rangle )\rightarrow
|g_{1}\rangle ,  \label{rot-ef}
\end{eqnarray}
and, therefore, 
\begin{equation}
|\psi \rangle _{A_1-C}=\frac{1}{\sqrt{2}}\{|f_{1}\rangle |0\rangle +
|g_{1}\rangle |1\rangle \}.
\end{equation}

Now, we let an atom $A_2$, which is prepared in the state 
\begin{equation}
| \psi \rangle _{A_2}=\frac{1}{\sqrt{2}}(| f_{2}\rangle +| g_{2}\rangle ),
\end{equation}
interact with the cavity. Taking into account (\ref{U-fg}), after the atom
has passed through it, we get 
\begin{equation}
| \psi \rangle _{A_1-A_2-C}=\frac{1}{2}\Biggl[| f_{1}\rangle \Bigl(| f_{2}\rangle +|g_{2}\rangle \Bigr)|0\rangle +| g_{1}\rangle \Bigl(-| f_{2}\rangle +| g_{2}\rangle \Bigr)|1\rangle \Biggr],
\end{equation}
Then, atom $A_2$ enters a Ramsey cavity $R_3$, where the atomic states are
rotated in the same way of (\ref{rot-ef}), so that we get 
\begin{equation}
|\psi \rangle _{A_1-A_2-C}=\frac{1}{\sqrt{2}}(| f_{1}\rangle | f_{2}\rangle
|0\rangle +| g_{1}\rangle | g_{2}\rangle |1\rangle ).
\end{equation}

In order to disentangle the atomic states of the field state, we send an
atom $A_3$, with the transition $|f_{k}\rangle \rightleftharpoons
|e_{k}\rangle $ resonant with the cavity, through $C$. If $A_3$ is sent in
the lower state $|f_{3}\rangle $, for an interaction time such that $g\tau=\pi /2$, we know that the state $|f_{3}\rangle |0\rangle $ does not evolve, whereas the state $|f_{3}\rangle |1\rangle $ evolves to $-i|e_{3}\rangle |0\rangle $. Then we get 
\begin{equation}
| \psi \rangle _{A_1-A_2-A_3-C}=\frac{1}{\sqrt{2}}\Bigl(| f_{1}\rangle |
f_{2}\rangle |f_{3}\rangle -i| g_{1}\rangle | g_{2}\rangle |e_{3}\rangle 
\Bigl)\otimes|0\rangle ,
\end{equation}
At this point, in comparison with the previous section, we have to perform an extra rotation. We let atom $A_3$ enter a Ramsey cavity $R_4$, where the
atomic states are rotated according to (\ref{rot_0}), which corresponds to
the transformation 
\begin{equation}
|e_{3}\rangle \rightarrow \frac{1}{\sqrt{2}}(| e_{3}\rangle +i| f_{3}\rangle
) ~~\mbox{and}~~ |f_{3}\rangle \rightarrow \frac{1}{\sqrt{2}}(i|
e_{3}\rangle +| f_{3}\rangle ),
\end{equation}
resulting in 
\begin{equation}
| \psi \rangle _{A_1-A_2-A_3}=\frac{1}{2}\Bigl\{| f_{1}\rangle | f_{2}\rangle
(~i| e_{3}\rangle +| f_{3}\rangle ~)-i| g_{1}\rangle | g_{2}\rangle (~|
e_{3}\rangle +i| f_{3}\rangle ~) \Bigr\}.
\end{equation}
Now, we have two possibilities of measurement concerning atom $A_3$: we may
detect it in state $|f_{3}\rangle $ and then we get the Bell state 
\begin{equation}
| \Phi ^{+}\rangle _{A_1-A_2}=\frac{1}{\sqrt{2}}(| f_{1}\rangle |
f_{2}\rangle +| g_{1}\rangle | g_{2}\rangle ),  \label{PHI+Casc}
\end{equation}
or in the state $|e_{3}\rangle $ and we have 
\begin{equation}
| \Phi ^{-}\rangle _{A_1-A_2}=\frac{1}{\sqrt{2}}(| f_{1}\rangle |
f_{2}\rangle -| g_{1}\rangle | g_{2}\rangle ).  \label{PHI-Casc}
\end{equation}
Now, if we apply an extra rotation on the states of atom $A_2$ in (\ref%
{PHI+Casc}) in a Ramsey cavity $R_5$, corresponding to the operator 
\begin{equation}
R_5=| f_{2}\rangle \langle g_{2}|+| g_{2}\rangle \langle f_{2}|,  \label{RC}
\end{equation}
we get 
\begin{equation}
| \Psi ^{+}\rangle _{A_1-A_2}=\frac{1}{\sqrt{2}}(| f_{1}\rangle |
g_{2}\rangle +| g_{1}\rangle | f_{2}\rangle ).  \label{PSI+Casc}
\end{equation}
Likewise, if we apply (\ref{RC}) to (\ref{PHI-Casc}), we get 
\begin{equation}
| \Psi ^{-}\rangle _{A_1-A_2}=\frac{1}{\sqrt{2}}(| f_{1}\rangle |
g_{2}\rangle -| g_{1}\rangle | f_{2}\rangle ).  \label{PSI-Casc}
\end{equation}
Therefore, as done before, we have obtained the Bell basis \cite{Nielsen,
BELLbasis} for the system $A_1-A_2$ given by Eqs (\ref{PHI+Casc}), (\ref%
{PHI-Casc}), (\ref{PSI+Casc}) and (\ref{PSI-Casc}).

In these two sections we have presented schemes to prepare atomic Bell
states involving three-level cascade atoms with two different cavity
configurations. In the former case one deals with a cavity prepared in a
coherent state $|-\alpha\rangle$, which is a feasible experimental task, but
during the process one has to inject another coherent state inside
the cavity and the final preparation of a Bell state depends also of a probabilistic detection of the state $|e_3\rangle$.
In the latter case one does not have to inject any field inside the cavity
and the process is not probabilistic. However, one has to
perform an extra rotation in the atomic states.

\subsection{Lambda atomic system and cavity in a coherent state}

Let us now show how we can get the Bell states making use of three-level
lambda atoms interacting with a cavity field. Consider a three-level lambda
atom (see Fig. 2) interacting with the electromagnetic field inside a cavity 
$C$. The states of the atom, $|a\rangle ,$ $ |b\rangle $ and $|c\rangle $
are so that the $|a\rangle \rightleftharpoons |c\rangle $ and $|a\rangle
\rightleftharpoons |b\rangle $ transitions are in the far off resonance
interaction limit. The time evolution operator for the atom-field
interaction $U(\tau)$ is given by \cite{Knight} 
\begin{equation}
U=-e^{i\varphi a^{\dagger }a}|a\rangle \langle a|+\frac{1}{2}(e^{i\varphi
a^{\dagger }a}+1)|b\rangle \langle b|+\frac{1}{2}(e^{i\varphi a^{\dagger
}a}-1)|b\rangle \langle c|\ +\frac{1}{2}(e^{i\varphi a^{\dagger
}a}-1)|c\rangle \langle b|+\frac{1}{2}(e^{i\varphi a^{\dagger
}a}+1)|c\rangle \langle c|,  \label{U1lambda}
\end{equation}
where $a$ $(a^{\dagger })$ is the annihilation (creation) operator for the
field in cavity $C$, $\varphi =2g^{2}\tau /$ $\Delta $, \ $g$ is the
coupling constant, $\Delta =\omega _{a}-\omega _{b}-\omega =\omega
_{a}-\omega _{c}-\omega $ is the detuning where \ $\omega _{a}$, $\omega
_{b} $ and $\omega _{c}$\ are the frequency of the upper level and \ of the
two degenerate lower levels respectively and $\omega $ is the cavity field
frequency and $\tau $ is the atom-field interaction time. For $\varphi =\pi $, we get 
\begin{equation}
U=-\exp \left( i\pi a^{\dagger }a\right) |a\rangle \langle a|+\Pi
_{+}|b\rangle \langle b|+\Pi _{-}|b\rangle \langle c|\ +\Pi _{-}|c\rangle
\langle b|+\Pi _{+}|c\rangle \langle c|,  \label{UlambdaPi}
\end{equation}
where 
\begin{eqnarray}
\Pi _{+} &=&\frac{1}{2}(e^{i\pi a^{\dagger }a}+1),  \nonumber \\
\Pi _{-} &=&\frac{1}{2}(e^{i\pi a^{\dagger }a}-1).  \label{pi+-}
\end{eqnarray}
At this point, it is worth to define the non-normalized even and odd
coherent states \cite{EvenOddCS} 
\begin{equation}
|\pm\rangle _{CS} =|\alpha \rangle \pm |-\alpha \rangle ,  \label{+-}
\end{equation}
such that $_{CS}\langle \pm | \pm \rangle_{CS} =2\left( 1\pm e^{-2| \alpha |
^{2}}\right) $ \ and $_{CS}\langle +| -\rangle_{CS} =0$, where $CS$ refers to coherent state. In the following calculation we shall use the relations 
\begin{eqnarray}
\Pi _{+}|+\rangle_{CS} =~|+\rangle_{CS} &\mbox{and}& \Pi _{+}|-\rangle_{CS} =0,  \nonumber
\\
\Pi _{-}|-\rangle_{CS} =-|-\rangle_{CS} &\mbox{and}& \Pi _{-}|+\rangle_{CS} =0,
\end{eqnarray}
which are easily obtained from Eqs. (\ref{pi+-}) and (\ref{+-}), using $
e^{za^{\dagger }a}|\alpha \rangle =|e^{z}\alpha \rangle $.

Let us prepare the cavity $C$ in the coherent state $|\alpha \rangle $ and
consider the atom $A_1$ in the state $|\psi \rangle_{A_1} = |b_{1}\rangle$.
The initial state of the atom-cavity system is given by 
\begin{equation}
|\psi \rangle _{A_1-C} =|b_{1}\rangle |\alpha \rangle =|b_{1}\rangle 
\frac{1}{2}[|+\rangle_{CS} +|-\rangle_{CS} ].
\end{equation}
We now let atom $A_1$ fly through the cavity $C$. The state of the system
evolves according to the time evolution operator Eq. (\ref{UlambdaPi})
yielding 
\begin{equation}
|\psi \rangle _{A_1-C} = \frac{1}{2}\{|b_{1}\rangle |+\rangle_{CS} -|c_{1}\rangle |-\rangle_{CS} \}.
\end{equation}

Consider another three-level lambda atom $A_2$ prepared initially in the
state $|b_{2}\rangle $, which is going to pass through the cavity. Now, as
initial state of the system, we have 
\begin{equation}
|\psi \rangle _{A_1-A_2-C}=\frac{1}{2}\{|b_{1}\rangle |+\rangle_{CS}
-|c_{1}\rangle |-\rangle_{CS} \}|b_{2}\rangle .
\end{equation}
After this second atom has passed through the cavity, the system evolves to 
\begin{equation}
|\psi \rangle _{A_1-A_2-C} =\frac{1}{2}\{|b_{1}\rangle |b_{2}\rangle
|+\rangle_{CS} +|c_{1}\rangle |c_{2}\rangle |-\rangle_{CS} \}.  \label{LBPSIA_1A_2C}
\end{equation}
Now, we inject a coherent state $|\alpha \rangle $ in the cavity, that is,
we make use of $D(\beta )|\alpha \rangle =|\alpha +\beta \rangle $, and we
get 
\begin{eqnarray}
|\psi \rangle _{A_1-A_2-C} &=&\frac{1}{2}\Bigl\{~|b_{1}\rangle |b_{2}\rangle 
\Bigl(~|2\alpha \rangle +|0\rangle ~\Bigr) + |c_{1}\rangle |c_{2}\rangle 
\Bigl(~|2\alpha \rangle -|0\rangle ~\Bigr)~\Bigr\}  \nonumber \\
&=&\frac{1}{2}\Bigl\{~\Bigl(~|b_{1}\rangle |b_{2}\rangle +|c_{1}\rangle
|c_{2}\rangle ~\Bigr)|2\alpha \rangle +\Bigl(~|b_{1}\rangle |b_{2}\rangle
-|c_{1}\rangle |c_{2}\rangle ~\Bigr)|0\rangle ~\Bigr\}.
\end{eqnarray}
Again, in order to disentangle the atomic states of the cavity field state,
we follow the same that was done in the subsection (\ref{casc-coer}), i.e.,
we send a two-level atom $A_3,$ resonant with the cavity, in its lower state 
$|f_{3}\rangle $ ($|f_{3}\rangle $ and $|e_{3}\rangle $ being the lower and
upper levels respectively) through $C$. Then we get 
\begin{equation}
| \psi \rangle _{A_1-A_2-A_3-C}=\frac{1}{2}\Bigl\{~\Bigl(~|b_{1}\rangle
|b_{2}\rangle + |c_{1}\rangle |c_{2}\rangle~\Bigr)\Bigl(~|e_{3}\rangle |\chi_e\rangle +|f_{3}\rangle |\chi _{f}\rangle~\Bigr) + \Bigl(~|b_{1}\rangle
|b_{2}\rangle -|c_{1}\rangle |c_{2}\rangle~\Bigr) |f_{3}\rangle |0\rangle ~
\Bigr\}.
\end{equation}
If we detect atom $A_3$ in state $|e_{3}\rangle$, then we finally get a Bell
state involving the atoms $A_1$ and $A_2$ 
\begin{equation}
| \Phi ^{+}\rangle _{A_1-A_2}=\frac{1}{\sqrt{2}}(|b_{1}\rangle |b_{2}\rangle
+|c_{1}\rangle |c_{2}\rangle ).  \label{LBBellPHI+}
\end{equation}

As mentioned in subsection (\ref{casc-coer}), the detection of atom $A_3$ in state $|e_{3}\rangle$ is probabilistic. Thus, in the above disentanglement process one has
to choose a coherent field with a photon-number distribution having a sharp
peak at the average photon number $\langle n\rangle =|\alpha |^{2}$ so that,
to a good approximation, we have almost $100\%$ of probability of detecting
the atom $A_3$ in the state $|e_{3}\rangle$ and, consequently, disentangling
successfully the atomic and field states, as it is desired.

The other Bell states can be obtained following the same procedure described
in the first subsection. Hence, if we start from (\ref{LBPSIA_1A_2C})
injecting a coherent state $|-\alpha \rangle $ in the cavity and detect $%
|e_{3}\rangle ,$ we get 
\begin{equation}
| \Phi ^{-}\rangle _{A_1-A_2}=\frac{1}{\sqrt{2}}(|b_{1}\rangle |b_{2}\rangle
-|c_{1}\rangle |c_{2}\rangle ).  \label{LBBellPHI-}
\end{equation}
Now, if we apply on the state (\ref{LBBellPHI+}) a rotation $R_1$ given by 
\begin{equation}
R_1 = | c_{2}\rangle \langle b_{2}|-| b_{2}\rangle \langle c_{2}|,
\label{R1}
\end{equation}
we get 
\begin{equation}
| \Psi ^{-}\rangle _{A_1-A_2}=\frac{1}{\sqrt{2}}(|b_{1}\rangle |c_{2}\rangle
-|c_{1}\rangle |b_{2}\rangle ),  \label{LBBellPSI-}
\end{equation}
and if we apply (\ref{R1}) to the state (\ref{LBBellPSI-}) we get 
\begin{equation}
| \Psi ^{+}\rangle _{A_1-A_2}=\frac{1}{\sqrt{2}}(|b_{1}\rangle |c_{2}\rangle
+|c_{1}\rangle |b_{2}\rangle ).  \label{LBBellPSI+}
\end{equation}
In the case of lambda atoms, the rotation of the atomic states (\ref{R1}) is
discussed in reference \cite{GHZLambdaat}.

Observe that, in comparison with the cascade atomic configuration, here one
needs only one rotation in a Ramsey cavity to obtain a Bell state.
Therefore, although the lambda atomic system is a more delicate
configuration to deal with because it involves a two-photon transition (two-photon Raman transition between the two degenerate atomic levels) contrary to the cascade case, which involves a one-photon transition, this configuration has the advantage of needing only one rotation to prepare the Bell state.

\subsection{Lambda atomic system and cavity in a superposition of Fock states}

Following the same steps of the last subsection, consider an atom $A_1$, in the state $|\psi \rangle _{A_1}=|b_{1}\rangle $, interacting with the cavity $C$ now prepared in the state (\ref{PsiC}).
Taking into account (\ref{UlambdaPi}), the state of the system atom-cavity
evolves to 
\begin{equation}
|\psi \rangle _{A_1-C}=\frac{1}{\sqrt{2}}(|b_{1}\rangle |0\rangle
-|c_{1}\rangle |1\rangle ).
\end{equation}
Consider now another three-level lambda atom $A_2$ prepared initially in the
state $|b_{2}\rangle $, which is going to pass through the cavity. After
this second atom has passed through the cavity, the system evolves to 
\begin{equation}
|\psi \rangle _{A_1-A_2-C}=\frac{1}{\sqrt{2}}(|b_{1}\rangle |b_{2}\rangle
|0\rangle +|c_{1}\rangle |c_{2}\rangle |1\rangle ).  \label{LBPSIA_1A_2C}
\end{equation}
In order to disentangle the atomic states of the cavity field state, as before, we now send a two-level atom $A_3,$ resonant with the cavity, with $|f_{3}\rangle $ and $|e_{3}\rangle $ being the lower and upper levels respectively, through $ C$. If $A_3$ is sent in the lower state $|f_{3}\rangle $, for $g\tau=\pi /2$, the state $|f_{3}\rangle |0\rangle $ does not evolve, whereas $|f_{3}\rangle |1\rangle $ evolves to $-i|e_{3}\rangle |0\rangle$. Then we get 
\begin{equation}
|\psi (\tau )\rangle _{A_1-A_2-A_3-C}=\frac{1}{\sqrt{2}}(|b_{1}\rangle
|b_{2}\rangle |f_{3}\rangle -i|c_{1}\rangle |c_{2}\rangle |e_{3}\rangle
)|0\rangle .
\end{equation}
Now we let atom $A_3$ to enter a Ramsey cavity $R_1$ where the atomic states
are rotated according to the rotation matrix (\ref{rot_0}), that is, 
\begin{eqnarray}
&&| e_{3}\rangle \rightarrow \frac{1}{\sqrt{2}}(| e_{3}\rangle +i|
f_{3}\rangle ),  \nonumber \\
&&| f_{3}\rangle \rightarrow \frac{1}{\sqrt{2}}(i| e_{3}\rangle +|
f_{3}\rangle ),
\end{eqnarray}
and we get 
\begin{equation}
|\psi \rangle _{A_1-A_2-A_3-C}=\frac{1}{2}\Bigl\{~|b_{1}\rangle |b_{2}\rangle
|f_{3}\rangle \Bigl(~i| e_{3}\rangle +| f_{3}\rangle ~\Bigr) -
i|c_{1}\rangle |c_{2}\rangle \Bigl(~| e_{3}\rangle +i| f_{3}\rangle ~\Bigr)
\Bigr\}|0\rangle .
\end{equation}
Now, if we detect $|f_{3}\rangle $, we get the Bell state 
\begin{equation}
| \Phi ^{+}\rangle _{A_1-A_2}=\frac{1}{\sqrt{2}}(|b_{1}\rangle |b_{2}\rangle
+|c_{1}\rangle |c_{2}\rangle ),  \label{PHI+Lamb}
\end{equation}
and, if we detect $|e_{3}\rangle $, we have 
\begin{equation}
| \Phi ^{-}\rangle _{A_1-A_2}=\frac{1}{\sqrt{2}}(| b_{1}\rangle |
b_{2}\rangle -| c_{1}\rangle | c_{2}\rangle ).  \label{PHI-Lamb}
\end{equation}
Now, if we apply an extra rotation on the states of atom $A_2$ in (\ref%
{PHI+Lamb}) in a Ramsey cavity $R_2$, according to the rotation matrix 
\begin{equation}
R_2=|b_{2}\rangle \langle c_{2}|+| c_{2}\rangle \langle b_{2}|,  \label{R}
\end{equation}
we get 
\begin{equation}
| \Psi ^{+}\rangle _{A_1-A_2}=\frac{1}{\sqrt{2}}(| b_{1}\rangle |
c_{2}\rangle +| c_{1}\rangle | b_{2}\rangle ),  \label{PSI+Lamb}
\end{equation}
and applying (\ref{R}) to (\ref{PHI-Casc}), it yields 
\begin{equation}
| \Psi ^{-}\rangle _{A_1-A_2}=\frac{1}{\sqrt{2}}(| b_{1}\rangle |
c_{2}\rangle -| c_{1}\rangle | b_{2}\rangle ).  \label{PSI-Lamb}
\end{equation}

\section{ENTANGLEMENT SWAPPING AND BELL STATE DETECTION}

Consider that we have a system, which consists of two pairs of entangled
atoms, whose state is 
\begin{equation}
|\Psi\rangle_{A_1-A_2-A_3-A_4}= |\Psi^-\rangle_{A_1-A_2} \otimes
|\Psi^-\rangle_{A_3-A_4}.
\end{equation}
The above state can be rewritten as 
\begin{eqnarray}
|\Psi\rangle_{A_1-A_2-A_3-A_4}&=& {\frac{1}{2}} \Bigl( |\Psi^+%
\rangle_{A_1-A_4}|\Psi^+\rangle_{A_2-A_3} +
|\Psi^-\rangle_{A_1-A_4}|\Psi^-\rangle_{A_2-A_3} +  \nonumber \\
& & ~~~~|\Phi^+\rangle_{A_1-A_4}|\Phi^+\rangle_{A_2-A_3} +
|\Phi^-\rangle_{A_1-A_4}|\Phi^-\rangle_{A_2-A_3} \Bigr) ,  \label{PSI1234}
\end{eqnarray}
where for the cascade system the Bell states are given by (\ref{EPRPHI+}), (
\ref{EPRPHI-}), (\ref{EPRPSI-}) and (\ref{EPRPSI+}) and for the lambda
system the Bell states are given by (\ref{LBBellPHI+}), (\ref{LBBellPHI-}), (\ref{LBBellPSI-}) and (\ref{LBBellPSI+}). Although atoms $A_1$ and $A_4$ has never interacted, we can entangle them if we measure properly the state
of atoms $A_2$ and $A_3$. This is what Eq. (\ref{PSI1234}) tell us. In the
next subsections we discuss how to perform the measurements in order to
project the state of $A_1$ and $A_4$ onto any of the Bell states.

\subsection{Bell state detection -- cascade atomic system and cavity in a
coherent state}

Let us assume we have a cavity prepared in a coherent state $|\alpha\rangle$. Notice that, according to the time evolution operator (\ref{U-fg}), if we
send atoms $A_2$ and $A_3$ through $C$ in the state (\ref{PHI+Casc}) or (\ref%
{PHI-Casc}), we get 
\begin{equation}
| \Phi ^{\pm }\rangle _{A_2-A_3}|\alpha \rangle \longrightarrow | \Phi ^{\pm
}\rangle _{A_2-A_3}|\alpha\rangle,
\end{equation}
and, if we send atoms $A_2$ and $A_3$ through $C$ in the state (\ref%
{PSI+Casc}) or (\ref{PSI-Casc}), we get 
\begin{equation}
| \Psi ^{\pm }\rangle _{A_2-A_3}|\alpha\rangle \longrightarrow | \Psi ^{\pm
}\rangle _{A_2-A_3}|-\alpha \rangle,
\end{equation}
Therefore, considering (\ref{PSI1234}), after atoms $A_2$ and $A_3$ pass
through the cavity, we have 
\begin{eqnarray}
|\Psi\rangle_{A_1-A_2-A_3-A_4-C}&=& {\frac{1}{2}} \Bigl( |\Psi^+%
\rangle_{A_1-A_4}|\Psi^+\rangle_{A_2-A_3}|-\alpha\rangle +
|\Psi^-\rangle_{A_1-A_4}|\Psi^-\rangle_{A_2-A_3}|-\alpha\rangle +  \nonumber
\\
& & ~~~~|\Phi^+\rangle_{A_1-A_4}|\Phi^+\rangle_{A_2-A_3}|\alpha\rangle +
|\Phi^-\rangle_{A_1-A_4}|\Phi^-\rangle_{A_2-A_3}|\alpha\rangle \Bigr).
\label{PSI1234C}
\end{eqnarray}
Now, we inject $|\alpha\rangle$ in the cavity $C$ and it yields 
\begin{eqnarray}
|\Psi\rangle_{A_1-A_2-A_3-A_4-C}&=& {\frac{1}{2}} \Bigl( |\Psi^+%
\rangle_{A_1-A_4}|\Psi^+\rangle_{A_2-A_3}|0\rangle +
|\Psi^-\rangle_{A_1-A_4}|\Psi^-\rangle_{A_2-A_3}|0\rangle +  \nonumber \\
& &~~~~ |\Phi^+\rangle_{A_1-A_4}|\Phi^+\rangle_{A_2-A_3}|2\alpha\rangle +
|\Phi^-\rangle_{A_1-A_4}|\Phi^-\rangle_{A_2-A_3}|2\alpha\rangle \Bigr) ,
\end{eqnarray}
As done in section (2.1), in order to disentangle the atomic states of the
cavity field state, we send a two-level atom $A_5,$ resonant with the
cavity, with $|f_{5}\rangle $ and $|e_{5}\rangle $ being the lower and upper
levels respectively, through $C$. If $A_5$ is sent in the lower state $%
|f_{5}\rangle$, after we detect $A_5$ in $|e_{5}\rangle$, we get 
\begin{equation}
|\Psi\rangle_{A_1-A_2-A_3-A_4} = {\frac{1}{\sqrt{2}}} \Bigl(%
|\Phi^+\rangle_{A_1-A_4}|\Phi^+\rangle_{A_2-A_3} +
|\Phi^-\rangle_{A_1-A_4}|\Phi^-\rangle_{A_2-A_3} \Bigr).
\end{equation}
Now, we have to distinguish $| \Phi ^{+}\rangle _{A_2-A_3}$ from $|\Phi
^{-}\rangle _{A_2-A_3}$. In order to do this we notice that, defining 
\begin{equation}
\Sigma _{x}=\sigma _{x}^{2}\sigma _{x}^{3},
\end{equation}
where 
\begin{equation}
\sigma _{x}^{k}=| f_{k}\rangle \langle g_{k}| +| g_{k}\rangle \langle f_{k}|
,
\end{equation}
we have 
\begin{equation}
\Sigma _{x} |\Phi ^{\pm }\rangle _{A_2-A_3}=\pm | \Phi ^{\pm }\rangle
_{A_2-A_3}.  \label{AVSIGMAx}
\end{equation}
Therefore, we can distinguish between $|\Phi ^{+}\rangle _{A_2-A_3}$ and $|
\Phi^{-}\rangle _{A_2-A_3}$ performing measurements of $\Sigma _{x}$. In
order to do so, we proceed as follows. We make use of 
\begin{equation}
K_{k}=\frac{1}{\sqrt{2}}(| f_{k}\rangle \langle f_{k}| -| f_{k}\rangle
\langle g_{k}| +| g_{k}\rangle \langle f_{k}| +| g_{k}\rangle \langle g_{k}|
),  \label{KkEPR}
\end{equation}
to gradually unravel the Bell states. The eigenvectors of the operators $
\sigma _{x}^{k}$ are 
\begin{equation}
|\psi _{x}^{k},\pm \rangle =\frac{1}{\sqrt{2}}(| f_{k}\rangle \pm |
g_{k}\rangle ),  \label{PSIxEPR}
\end{equation}
and we can rewrite the Bell states as 
\begin{equation}
|\Phi ^{\pm }\rangle _{A_2-A_3}=\frac{1}{2}\Biggl[|\psi _{x}^{2},+\rangle 
\Bigl(| f_{3}\rangle \pm | g_{3}\rangle \Bigr) +|\psi _{x}^{2},-\rangle 
\Bigl(|f_{3}\rangle \mp | g_{3}\rangle \Bigr)\Biggr],  \label{EPRPHIx}
\end{equation}
Let us take, for instance, (\ref{PHI+Casc}): 
\begin{equation}
| \Phi ^{+}\rangle _{A_2-A_3}=\frac{1}{\sqrt{2}}(| f_{2}\rangle |
f_{3}\rangle +| g_{2}\rangle | g_{3}\rangle ).
\end{equation}
Applying $K_{2}$ to this state we have 
\begin{equation}
K_{2}| \Phi ^{+}\rangle _{A_2-A_3}=\frac{1}{2}\Biggl\{|f_{2}\rangle \Bigl(|
f_{3}\rangle -| g_{3}\rangle \Bigr)+|g_{2}\rangle \Bigl(| f_{3}\rangle +|
g_{3}\rangle \Bigr)\Biggr\}.  \label{EPRK1PHI23}
\end{equation}
Now, we compare (\ref{EPRK1PHI23}) and (\ref{EPRPHIx}). We see that the
rotation by $K_{2}$ followed by the detection of $|g_{2}\rangle $
corresponds to the detection of the state $|\psi _{x}^{2},+\rangle $
whose eigenvalue of $\sigma _{x}^{2}$ is $+1$. After we detect $
|g_{2}\rangle $, we get 
\begin{equation}
| \psi \rangle _{A_3}=\frac{1}{\sqrt{2}}(| f_{3}\rangle +| g_{3}\rangle ),
\end{equation}
that is, we have got 
\begin{equation}
| \psi \rangle _{A_3}=|\psi _{x}^{3},+\rangle .  \label{EPRPSI2x}
\end{equation}
If we apply (\ref{KkEPR}) for $k=3$ to the state (\ref{EPRPSI2x}) we get 
\begin{equation}
K_{3}| \psi \rangle _{A_3}=|g_{3}\rangle .
\end{equation}
We see that the rotation by $K_{3}$ followed by the detection of $
|g_{3}\rangle $ corresponds to the detection of the state $|\psi
_{x}^{3},+\rangle $ whose eigenvalue of $\sigma _{x}^{3}$ is $+1$.
Consequently, after this proceeding, the atoms $A_1$ and $A_4$ are collapsed
in the entangled state $|\Phi^+\rangle_{A_1-A_4}$. Similarly, we can measure
the eigenvalue $-1$ of $\Sigma_x$ and, consequently, make the atoms $A_1$
and $A_4$ to collapse in the entangled state $|\Phi^-\rangle_{A_1-A_4}$.

If we had injected $|-\alpha\rangle$ in $C$, we would get 
\begin{eqnarray}
|\Psi\rangle_{A_1-A_2-A_3-A_4-C}&=& {\frac{1}{2}} \Bigl( |\Psi^+%
\rangle_{A_1-A_4}|\Psi^+\rangle_{A_2-A_3}|-2\alpha\rangle +
|\Psi^-\rangle_{A_1-A_4}|\Psi^-\rangle_{A_2-A_3}|-2\alpha\rangle +  \nonumber
\\
& &~~~~ |\Phi^+\rangle_{A_1-A_4}|\Phi^+\rangle_{A_2-A_3}|0\rangle +
|\Phi^-\rangle_{A_1-A_4}|\Phi^-\rangle_{A_2-A_3}|0\rangle \Bigr)
\end{eqnarray}
and, as above, after we send atom $A_5$ through $C$ and detect $|e_5\rangle$
, we get 
\begin{equation}
|\Psi\rangle_{A_1-A_2-A_3-A_4}= {\frac{1}{2}} \Bigl( |\Psi^+%
\rangle_{A_1-A_4}|\Psi^+\rangle_{A_2-A_3} +
|\Psi^-\rangle_{A_1-A_4}|\Psi^-\rangle_{A_2-A_3} \Bigr).
\end{equation}
Now we have to distinguish $| \Psi ^{+}\rangle _{A_2-A_3}$ from $|\Psi
^{-}\rangle _{A_2-A_3}$. In order to do this, as above, we measure the
eigenvalues of the operator $\Sigma _{x}$ 
\begin{equation}
\Sigma _{x} |\Psi ^{\pm }\rangle _{A_2-A_3}=\pm | \Psi ^{\pm }\rangle
_{A_2-A_3}.
\end{equation}
After we have measured the eigenvalue $+1$ of $\Sigma_x$, the atoms $A_1$
and $A_4$ are collapsed in the entangled state $|\Psi^+\rangle_{A_1-A_4}$,
and if we have measured the eigenvalue $-1$ of $\Sigma_x$, the atoms $A_1$
and $A_4$ would be collapsed in the entangled state $|\Psi^-\rangle_{A_1-A_4}$.

Summarizing, we have the following possible proceedings which results in one
of the four Bell states involving atoms $A_1$ and $A_4$, which are presented
in the table below: 
\begin{eqnarray}
& &(\mbox{injection of $|\alpha\rangle$})(K_{2}, |g_{2}\rangle )(K_{3},|
g_{3}\rangle )\longleftrightarrow |\psi _{x}^{2},+\rangle |\psi
_{x}^{3},+\rangle \Longrightarrow |\Phi^+\rangle_{A_1-A_4}  \nonumber \\
& &(\mbox{injection of $|\alpha\rangle$})(K_{2}, |g_{2}\rangle )(K_{3},|
f_{3}\rangle )\longleftrightarrow |\psi _{x}^{2},+\rangle |\psi
_{x}^{3},-\rangle \Longrightarrow |\Phi^-\rangle_{A_1-A_4}  \nonumber \\
& &(\mbox{injection of $|-\alpha\rangle$})(K_{2}, |f_{2}\rangle )(K_{3},|
f_{3}\rangle )\longleftrightarrow |\psi _{x}^{2},-\rangle |\psi
_{x}^{3},-\rangle \Longrightarrow |\Psi^+\rangle_{A_1-A_4}  \nonumber \\
& &(\mbox{injection of $|-\alpha\rangle$})(K_{2}, |g_{2}\rangle )(K_{3},|
f_{3}\rangle )\longleftrightarrow |\psi _{x}^{2},+\rangle |\psi
_{x}^{3},-\rangle \Longrightarrow |\Psi^-\rangle_{A_1-A_4}
\end{eqnarray}

\subsection{Bell state detection -- cascade atomic system and cavity in a
superposition of Fock states}

Now let us see how to distinguish the four states which form the Bell basis
in this case. Let us assume that we have a cavity $C$ prepared in the state (\ref{PsiC}). Observe that if we send atoms $A_2$ and $A_3$ through $C$ in the state (\ref{PHI+Casc}) or (\ref{PHI-Casc}), we get 
\begin{equation}
| \Phi ^{\pm }\rangle _{A_2-A_3}|+ \rangle _{FS} \longrightarrow | \Phi
^{\pm }\rangle _{A_2-A_3}|+ \rangle _{FS},
\end{equation}
and if we send atoms $A_2$ and $A_3$ through $C$ in the state (\ref{PSI+Casc}
) or (\ref{PSI-Casc}), we get 
\begin{equation}
| \Psi ^{\pm }\rangle _{A_2-A_3}|+ \rangle _{FS} \longrightarrow | \Psi
^{\pm }\rangle _{A_2-A_3}|-\rangle _{FS},
\end{equation}
where we have defined 
\begin{equation}
|-\rangle _{FS} = \frac{(|0\rangle -|1\rangle )}{\sqrt{2}}.
\end{equation}
Then, starting from (\ref{PSI1234}), after atoms $A_2$ and $A_3$ cross the
cavity, we have 
\begin{eqnarray}
|\Psi\rangle_{A_1-A_2-A_3-A_4-C}&=& {\frac{1}{2}} \Bigl( |\Psi^+%
\rangle_{A_1-A_4}|\Psi^+\rangle_{A_2-A_3}|-\rangle _{FS} +
|\Psi^-\rangle_{A_1-A_4}|\Psi^-\rangle_{A_2-A_3}|-\rangle _{FS} + 
\nonumber \\
& & |\Phi^+\rangle_{A_1-A_4}|\Phi^+\rangle_{A_2-A_3}|+ \rangle _{FS} +
|\Phi^-\rangle_{A_1-A_4}|\Phi^-\rangle_{A_2-A_3} |+ \rangle _{FS} \Bigr) %
,
\end{eqnarray}
Now, if an atom $A_5$, resonant with the cavity, passes through $C$ in the
state $|f_{5}\rangle $ ( $|f_{5}\rangle $ and $|e_{5}\rangle $ being the
lower and upper level, respectively), for $g\tau=\pi /2$, we have 
\begin{eqnarray}
|f_{5}\rangle |+ \rangle _{FS} &\longrightarrow &\frac{1}{\sqrt{2}}
(|f_{5}\rangle -i|e_{5}\rangle )|0\rangle ,  \nonumber \\
|f_{5}\rangle |-\rangle _{FS} &\longrightarrow &\frac{1}{\sqrt{2}}
(|f_{5}\rangle +i|e_{5}\rangle )|0\rangle .  \label{e5f5a}
\end{eqnarray}
Consider now that we send atom $A_5$ through a Ramsey cavity $R$, where the
states are rotated according to the rotation matrix (\ref{rot_0}), it yields 
\begin{eqnarray}
&&\frac{1}{\sqrt{2}}(|f_{5}\rangle - i | e_{5}\rangle)\rightarrow |
f_{5}\rangle ,  \nonumber \\
&&\frac{1}{\sqrt{2}}(|f_{5}\rangle + i | e_{5}\rangle )\rightarrow i|
e_{5}\rangle.  \label{e5f5b}
\end{eqnarray}
Therefore, after $A_5$ has passed through the cavity, we get 
\begin{eqnarray}
|\Psi\rangle_{A_1-A_2-A_3-A_4-A_5}&=& {\frac{1}{2}} \Bigl( %
|\Psi^+\rangle_{A_1-A_4}|\Psi^+\rangle_{A_2-A_3}i | e_{5}\rangle +
|\Psi^-\rangle_{A_1-A_4}|\Psi^-\rangle_{A_2-A_3}i | e_{5}\rangle +  \nonumber
\\
& & |\Phi^+\rangle_{A_1-A_4}|\Phi^+\rangle_{A_2-A_3}|f_{5}\rangle +
|\Phi^-\rangle_{A_1-A_4}|\Phi^-\rangle_{A_2-A_3} |f_{5}\rangle \Bigr) .
\end{eqnarray}
Now, if we detect $|e_5\rangle$, we get 
\begin{equation}
|\Psi\rangle_{A_1-A_2-A_3-A_4}= {\frac{1}{\sqrt{2}}} \Bigl( %
|\Psi^+\rangle_{A_1-A_4}|\Psi^+\rangle_{A_2-A_3} +
|\Psi^-\rangle_{A_1-A_4}|\Psi^-\rangle_{A_2-A_3} \Bigr) 
\end{equation}
and, as in the previous subsection, if we measure the eigenvalue $+1$ of $%
\Sigma_x$, the atoms $A_1$ and $A_4$ collapse in the state $%
|\Psi^+\rangle_{A_1-A_4}$, and if we measure the eigenvalue $-1$ of $\Sigma_x$%
, they collapse in the state $|\Psi^-\rangle_{A_1-A_4}$.

But, if we detect $| f_{5}\rangle$ instead of $|e_5\rangle$, we get 
\begin{equation}
|\Psi\rangle_{A_1-A_2-A_3-A_4}= {\frac{1}{\sqrt{2}}} \Bigl( %
|\Phi^+\rangle_{A_1-A_4}|\Phi^+\rangle_{A_2-A_3} +
|\Phi^-\rangle_{A_1-A_4}|\Phi^-\rangle_{A_2-A_3} \Bigr).
\end{equation}
And, finally, as in the previous section, by measuring the eigenvalue $+1$
of $\Sigma_x$, the atoms $A_1$ and $A_4$ collapse in the state $%
|\Phi^+\rangle_{A_1-A_4}$, and if we measure the eigenvalue $-1$ of $\Sigma_x$%
, they collapse in the state $|\Phi^-\rangle_{A_1-A_4}$.

Summarizing, we have the following possible proceedings which result in one
of the four Bell states involving atoms $A_1$ and $A_4$, which are presented
in the table below: 
\begin{eqnarray}
& &(\mbox{detection of $|f_5\rangle$})(K_{2}, |g_{2}\rangle )(K_{3},|
g_{3}\rangle )\longleftrightarrow |\psi _{x}^{2},+\rangle |\psi
_{x}^{3},+\rangle \Longrightarrow |\Phi^+\rangle_{A_1-A_4}  \nonumber \\
& &(\mbox{detection of $|f_5\rangle$})(K_{2}, |g_{2}\rangle )(K_{3},|
f_{3}\rangle )\longleftrightarrow |\psi _{x}^{2},+\rangle |\psi
_{x}^{3},-\rangle \Longrightarrow |\Phi^-\rangle_{A_1-A_4}  \nonumber \\
& &(\mbox{detection of $|e_5\rangle$})(K_{2}, |f_{2}\rangle )(K_{3},|
f_{3}\rangle )\longleftrightarrow |\psi _{x}^{2},-\rangle |\psi
_{x}^{3},-\rangle \Longrightarrow |\Psi^+\rangle_{A_1-A_4}  \nonumber \\
& &(\mbox{detection of $|e_5\rangle$})(K_{2}, |g_{2}\rangle )(K_{3},|
f_{3}\rangle )\longleftrightarrow |\psi _{x}^{2},+\rangle |\psi
_{x}^{3},-\rangle \Longrightarrow |\Psi^-\rangle_{A_1-A_4}
\end{eqnarray}

\subsection{Bell state detection -- lambda atomic system and cavity in a
coherent state}

Now, consider that we have a cavity prepared in a coherent state $%
|\alpha\rangle$. Notice that if we send atoms $A_2$ and $A_3$ through $C$ in
the state (\ref{LBBellPHI+}) or (\ref{LBBellPHI-}), we get 
\begin{equation}
| \Phi ^{\pm }\rangle _{A_2-A_3}|\alpha \rangle \longrightarrow | \Phi ^{\pm
}\rangle _{A_2-A_3}|\pm\alpha\rangle
\end{equation}
and, if we send atoms $A_2$ and $A_3$ through $C$ in the state (\ref%
{LBBellPSI+}) or (\ref{LBBellPSI-}), it yields 
\begin{equation}
| \Psi ^{\pm }\rangle _{A_2-A_3}|\alpha\rangle \longrightarrow | \Psi ^{\pm
}\rangle _{A_2-A_3}|\pm\alpha \rangle.
\end{equation}
Therefore, considering (\ref{PSI1234}), after atoms $A_2$ and $A_3$ pass
through the cavity, we have 
\begin{eqnarray}
|\Psi\rangle_{A_1-A_2-A_3-A_4-C}&=& {\frac{1}{2}} \Bigl( |\Psi^+%
\rangle_{A_1-A_4}|\Psi^+\rangle_{A_2-A_3}|\alpha\rangle +
|\Psi^-\rangle_{A_1-A_4}|\Psi^-\rangle_{A_2-A_3}|-\alpha\rangle +  \nonumber
\\
& & ~~~~|\Phi^+\rangle_{A_1-A_4}|\Phi^+\rangle_{A_2-A_3}|\alpha\rangle +
|\Phi^-\rangle_{A_1-A_4}|\Phi^-\rangle_{A_2-A_3}|-\alpha\rangle \Bigr) .
\label{PSI1234C}
\end{eqnarray}

Now we inject $|\alpha\rangle$ in the cavity $C$ and we have 
\begin{eqnarray}
|\Psi\rangle_{A_1-A_2-A_3-A_4-C}&=& {\frac{1}{2}} \Bigl( |\Psi^+%
\rangle_{A_1-A_4}|\Psi^+\rangle_{A_2-A_3}|2\alpha\rangle +
|\Psi^-\rangle_{A_1-A_4}|\Psi^-\rangle_{A_2-A_3}|0\rangle +  \nonumber \\
& & ~~~~|\Phi^+\rangle_{A_1-A_4}|\Phi^+\rangle_{A_2-A_3}|2\alpha\rangle +
|\Phi^-\rangle_{A_1-A_4}|\Phi^-\rangle_{A_2-A_3}|0\rangle \Bigr) .
\end{eqnarray}
As done in subsection (2.1), in order to disentangle the atomic states of the
cavity field state we send a two-level atom $A_5,$ resonant with the cavity,
with $|f_{5}\rangle $ and $|e_{5}\rangle $ being the lower and upper levels
respectively, through $C$. If $A_5$ is sent in the lower state $|f_{5}\rangle
$, after we detect $A_5$ in $|e_{5}\rangle$, we have 
\begin{equation}
|\Psi\rangle_{A_1-A_2-A_3-A_4} = {\frac{1}{2}} \Bigl( |\Psi^+%
\rangle_{A_1-A_4}|\Psi^+\rangle_{A_2-A_3} + |\Phi^+\rangle_{A_1-A_4}
|\Phi^+\rangle_{A_2-A_3} \Bigr) .
\end{equation}
Then, we detect the states of atoms $A_2$ and $A_3$. If we detect $%
|b_2\rangle$ and $|b_3\rangle$ or $|c_2\rangle$ and $|c_3\rangle$, the atoms 
$A_1$ and $A_4$ collapse to the state $|\Phi^+\rangle_{A_1-A_4}$, and if we
detect $|b_2\rangle$ and $|c_3\rangle$ or $|c_2\rangle$ and $|b_3\rangle$,
the atoms $A_1$ and $A_4$ collapse to the state $|\Psi^+\rangle_{A_1-A_4}$

On the other hand, if we inject $|-\alpha\rangle$ in the cavity $C$, we have 
\begin{eqnarray}
|\Psi\rangle_{A_1-A_2-A_3-A_4-C}&=& {\frac{1}{2}} \Bigl( |\Psi^+
\rangle_{A_1-A_4}|\Psi^+\rangle_{A_2-A_3}|0\rangle +
|\Psi^-\rangle_{A_1-A_4}|\Psi^-\rangle_{A_2-A_3}|-2\alpha\rangle +  \nonumber \\
& & ~~~~|\Phi^+\rangle_{A_1-A_4}|\Phi^+\rangle_{A_2-A_3}|0\rangle +
|\Phi^-\rangle_{A_1-A_4}|\Phi^-\rangle_{A_2-A_3}|-2\alpha\rangle \Bigr) .
\end{eqnarray}
As before, we use an auxiliary atom $A_5$ to disentangle the atomic states
of the cavity field state. If $A_5$ is sent in the lower state $|f_{5}\rangle
$ and detected $A_5$ in $|e_{5}\rangle$, we get 
\begin{equation}
|\Psi\rangle_{A_1-A_2-A_3-A_4}= {\frac{1}{2}} \Bigl( |\Psi^-%
\rangle_{A_1-A_4}|\Psi^-\rangle_{A_2-A_3} +
|\Phi^-\rangle_{A_1-A_4}|\Phi^-\rangle_{A_2-A_3} \Bigr) .
\end{equation}
And now, we detect the states of atoms $A_2$ and $A_3$. If we detect $%
|b_2\rangle$ and $|b_3\rangle$ or $|c_2\rangle$ and $|c_3\rangle$, the atoms 
$A_1$ and $A_4$ collapse to the state $|\Phi^-\rangle_{A_1-A_4}$, and if we
detect $|b_2\rangle$ and $|c_3\rangle$ or $|c_2\rangle$ and $|b_3\rangle$,
they collapse to the state $|\Psi^-\rangle_{A_1-A_4}$.

\subsection{Bell state detection -- lambda atomic system and cavity in a
superposition of Fock states}

Finally, consider a cavity prepared in the state (\ref{PsiC}). In this case,
observe that if we send atoms $A_2$ and $A_3$ through $C$ in the state (\ref%
{LBBellPHI+}) or (\ref{LBBellPHI-}), we get 
\begin{equation}
| \Phi ^{\pm }\rangle _{A_2-A_3}|\psi_+ \rangle_C \longrightarrow | \Phi
^{\pm }\rangle _{A_2-A_3}|\pm\rangle _{FS}
\end{equation}
and if we send atoms $A_2$ and $A_3$ through $C$ in the state (\ref%
{LBBellPSI+}) or (\ref{LBBellPSI-}), we have 
\begin{equation}
| \Psi ^{\pm }\rangle _{A_2-A_3}|\psi_+ \rangle_C \longrightarrow | \Psi
^{\pm }\rangle _{A_2-A_3}|\pm\rangle _{FS}.
\end{equation}
Therefore, considering (\ref{PSI1234}), after atoms $A_2$ and $A_3$ pass
through the cavity, it yields%
\begin{eqnarray}
|\Psi\rangle_{A_1-A_2-A_3-A_4-C}&=& {\frac{1}{2}} \Bigl( |\Psi^+
\rangle_{A_1-A_4}|\Psi^+\rangle_{A_2-A_3}|+\rangle_{FS} +
|\Psi^-\rangle_{A_1-A_4}|\Psi^-\rangle_{A_2-A_3}|-\rangle_{FS} +  \nonumber
\\
& &~~~~ |\Phi^+\rangle_{A_1-A_4}|\Phi^+\rangle_{A_2-A_3}|+\rangle_{FS} +
|\Phi^-\rangle_{A_1-A_4}|\Phi^-\rangle_{A_2-A_3}|-\rangle_{FS} \Bigr).
\end{eqnarray}

Again, as it was done in subsection 3.2, if we send an atom $A_5$, resonant
with the cavity, through $C$ in the state $|f_{5}\rangle $, for $
g\tau=\pi /2$, and through a Ramsey cavity, according to (\ref{e5f5a}) and (\ref{e5f5b}), we get 
\begin{eqnarray}
|\Psi\rangle_{A_1-A_2-A_3-A_4-A_5}&=& {\frac{1}{2}} \Bigl( %
|\Psi^+\rangle_{A_1-A_4}|\Psi^+\rangle_{A_2-A_3}|f_5\rangle +
|\Psi^-\rangle_{A_1-A_4}|\Psi^-\rangle_{A_2-A_3}i|e_5\rangle +  \nonumber \\
& &~~~~ |\Phi^+\rangle_{A_1-A_4}|\Phi^+\rangle_{A_2-A_3}|f_5\rangle +
|\Phi^-\rangle_{A_1-A_4}|\Phi^-\rangle_{A_2-A_3}i|e_5\rangle \Bigr).
\end{eqnarray}

Then, if we detect $A_5$ in the state $|f_{5}\rangle$, we have 
\begin{equation}
|\Psi\rangle_{A_1-A_2-A_3-A_4} = {\frac{1}{2}} \Bigl( |\Psi^+%
\rangle_{A_1-A_4}|\Psi^+\rangle_{A_2-A_3} + |\Phi^+\rangle_{A_1-A_4}
|\Phi^+\rangle_{A_2-A_3} \Bigr) .
\end{equation}
Now, we detect the states of atoms $A_2$ and $A_3$. If we detect $%
|b_2\rangle$ and $|b_3\rangle$ or $|c_2\rangle$ and $|c_3\rangle$, atoms $A_1
$ and $A_4$ collapse to the state $|\Phi^+\rangle_{A_1-A_4}$, and if we
detect $|b_2\rangle$ and $|c_3\rangle$ or $|c_2\rangle$ and $|b_3\rangle$,
atoms $A_1$ and $A_4$ collapse to the state $|\Psi^+\rangle_{A_1-A_4}$

Otherwise, if we detect $A_{5}$ in the state $|e_{5}\rangle $, we have 
\begin{equation}
|\Psi \rangle _{A_{1}-A_{2}-A_{3}-A_{4}}={\frac{1}{2}}\Bigl(|\Psi
^{-}\rangle _{A_{1}-A_{4}}|\Psi ^{-}\rangle _{A_{2}-A_{3}}+|\Phi ^{-}\rangle
_{A_{1}-A_{4}}|\Phi ^{-}\rangle _{A_{2}-A_{3}}\Bigr).
\end{equation}%
Then, we detect the states of atoms $A_{2}$ and $A_{3}$. If we detect $%
|b_{2}\rangle $ and $|b_{3}\rangle $ or $|c_{2}\rangle $ and $|c_{3}\rangle $%
, atoms $A_{1}$ and $A_{4}$ collapse to the state $|\Phi ^{-}\rangle
_{A_{1}-A_{4}}$, and if we detect $|b_{2}\rangle $ and $|c_{3}\rangle $ or $%
|c_{2}\rangle $ and $|b_{3}\rangle $, atoms $A_{1}$ and $A_{4}$ collapse to
the state $|\Psi ^{-}\rangle _{A_{1}-A_{4}}$.

\section{CONCLUSION}

Concluding, we have presented four schemes of entanglement swapping involving atoms in a cascade configuration or in a lambda configuration interacting with a cavity prepared in a coherent state or in a superposition of $|0\rangle$ and $|1\rangle$ Fock states. The advantage of using a cascade atomic configuration instead of a lambda one is that the atomic state detections are simpler. However, the process for the lambda configuration involves much less atomic state rotation than in the cascade configuration case. The advantage of using a cavity prepared in a superposition of Fock state is that the process is not probabilistic, while in the case in which the cavity is prepared in a coherent state the process is probabilistic, although the probability of the atomic state involved can be made close to $100\%$ for coherent state with a large mean photon number and a sharp peak distribution. On the other hand, the advantage of using a coherent state is that its preparation is relatively simple.

Finally, let us analyze the feasibility of the experimental implementation
of the above schemes of entanglement swapping. Considering Rydberg atoms of
principal quantum numbers 50 or 51, the radiative time is of the order of $
10^{-2}$ s and the coupling constant $g$ is of the order of $2\pi \times 25$
kHz \cite{Rat1, Rat2, Rat3} and the detuning $\Delta $ is of the order of $
2\pi \times 100$ kHz. Taking into account that, for the cascade configuration, $\varphi =g^{2}\tau /\Delta $ and, for the lambda configuration, $\varphi =2g^{2}\tau /\Delta $ , for $\varphi =\pi $ we have an interaction time $
\tau =8\times 10^{-5}$ s for the cascade configuration and $\tau =4\times
10^{-5}$ for the lambda configuration and we could, in principle, assume a
time of the order of $10^{-4}$ s to realize the entanglement swapping which is much shorter than the radiative time. We have to consider also the cavity decay time which in recent experiments, with niobium superconducting cavities at very low temperature and quality factors in the $10^{9}-10^{10}$  range, have a cavity energy damping time of the order of $10$ to $100$ ms,  and which could be larger than the required time to perform the entanglement swapping.
\newline

\textbf{Figure Captions} \newline

\textbf{Fig. 1 -} Energy states scheme of a three-level atom where $|e\rangle$ is the upper state with atomic frequency $\omega _{e}$, $\ |f\rangle $ is
the intermediate state with atomic frequency $\omega _{f}$, $|g\rangle $ is
the lower state with atomic frequency $\omega _{g}$ and $\omega $ is the
cavity field frequency and $\Delta =(\omega _{e}-\omega _{f})-\omega $ is
the detuning. The transition $\mid f\rangle \rightleftharpoons \mid e\rangle 
$ is far enough of resonance with the cavity central frequency such that
only virtual transitions occur between these levels (only these states
interact with field in cavity $C$). In addition we assume that the
transitions $|e\rangle \rightleftharpoons |g\rangle $ and $|f\rangle \rightleftharpoons |g\rangle $ are highly detuned from the cavity frequency so that there will be no coupling with the cavity field in $C$.\newline

\textbf{Fig. 2 -} Energy level scheme of the three-level lambda atom where $
|a\rangle $ is the upper state with atomic frequency $\omega _{a}$, $
|b\rangle $ \ and $|c\rangle $ are the lower states with atomic frequency $
\omega _{b}$ and $\omega _{c}$ respectively, $\omega $ is the cavity field frequency and $\Delta =\omega _{a}-\omega _{b}-\omega =\omega _{a}-\omega _{c}-\omega $ is
the detuning.\newline

\end{document}